\documentclass[a4paper,twoside]{article}
\pdfoutput=1
\usepackage{algorithm}
\usepackage[noend]{algorithmic} % can use noend to play with spacing
\usepackage{epsfig}
\usepackage{subfigure}
\usepackage{calc}
\usepackage{amssymb}
\usepackage{amstext}
\usepackage{amsmath}
\usepackage{multicol}
\usepackage{pslatex}

\usepackage{apalike}
\usepackage{fancyhdr}
\usepackage{insticc}
\usepackage{float}
\newcommand{\danielcut}[1]{}%
\newcommand{\daniel}[1]{\textcolor{red}{\textbf{#1}}}%
\newcommand{\kamel}[1]{}%
\usepackage[small]{caption}
\newcommand{\lxor}{\oplus}
\newtheorem{proposition}{Proposition}  %in llncs
\def\qed{\relax\ifmmode\hskip2em \Box\else\unskip\nobreak\hskip1em $\Box$\fi}
\newtheorem{corollary}{Corollary}  %in llncs
\newenvironment{proof}{Proof.}{\qed}
\usepackage{ifpdf}
\usepackage{color}
\ifpdf
    \providecommand{\myfig}[1]{#1.pdf}
\else
   
   \providecommand{\myfig}[1]{#1.eps}
\fi
\usepackage{url}
\begin{document}
\onecolumn\maketitle \normalsize \vfill

\section{\uppercase{Introduction}}

\noindent View materialization is presumably one of the most effective
technique to improve query performance of data warehouses. Materialized views
are physical structures that improve data access time by precomputing
intermediary results. 
Typical OLAP queries defined on data warehouses consist in selecting
and aggregating data with queries such as
 grouping sets (\textsc{group by} clauses)\danielcut{~\cite{graycube}}.  By precomputing
many plausible groupings, we can avoid aggregates over large tables.
However,
materializing views requires additional storage space and induces maintenance
overhead when refreshing the data warehouse.

One of the most important issues in data warehouse physical design is to select
an appropriate configuration of materialized views. Several heuristics and
methodologies were proposed for the  materialized view selection problem, which
is NP-hard~\cite{gup97sel}. Most of these techniques exploit cost models to
estimate the data access cost using materialized views, their maintenance and
storage cost. This cost estimation mostly depends on view-size  estimation.

%That is important to provide an accurate view-size estimation.

%  \noindent data warehouses have nearly large storage capacities but
% not everything can be materialized if only because of the processing time
% required. View-size estimation is generally important. estimate fast so that we
% can decide faster what to materialize. materialization strategies can assume
% accurate estimates. talk about GROUBPYS (define \textsc{group by} query).
%
%
% We need to balance
% estimation speed versus estimation accuracy.
% We argue that a method using 32~bytes is not necessarily
% superior to a method using 32KB when most data warehouse servers
% have several gigabytes of internal memory. Speed and accuracy and
% most important, as long as the memory usage is small. For example,
% repeatedly hashing the data is not an option.

%Cite some studies requiring assumptions for view-size estimation.

Several techniques have been proposed for view-size estimation:
some requiring assumptions about the data
distribution and others that are ``unassuming.'' %For example, assuming that the data
%is uniformly distributed, %Golfarelli~et~al.~\cite{gol98met}
%proposed to estimate the number of tuples in a given view by using Yao's~\cite{yao77app}
%and Cardenas'~\cite{car75ana}  formulas.
A common statistical assumption is uniformity~\cite{gol98met}, but any
skew in the data leads to an overestimate of the size of the view.
%and can only be used as  a crude estimation. Estimators that can adapt
% %to the distribution skew were also proposed~\cite{haas1995sbe}.
% We selected a competitive
% estimator based on a multifractal distribution~\cite{faloutsos1996msd} which is
% well adapted to the typically skewed distribution found in data warehousing.
Generally, while statistically assuming estimators are computed quickly, the most expensive
step being the random sampling, their error
can be large and it cannot be bounded a priori.

In this paper, we consider several state-of-the-art statistically unassuming
estimation techniques: \textsc{Gibbons-Tirthapura}~\cite{Gibbons2001},
probabilistic counting~\cite{flajolet1985pca}, and \textsc{LogLog}
probabilistic counting~\cite{durand2003lcl}. While relatively expensive,
unassuming estimators tend to provide a good accuracy. To our knowledge, this
is the first experimental comparisons of unassuming  view-size estimation
techniques in a data warehousing setting.

\section{\uppercase{Related Work}}\label{sec:RealtedWork}
\noindent Haas~et~al.~\cite{haas1995sbe}  estimate the view-size from the  histogram of a
sample: adaptively, they choose a different estimator based on the skew of
the distribution. Faloutsos~et~al.~\cite{faloutsos1996msd} obtain results nearly as accurate as
Haas et al., that is, an error of approximately 40\%, but they only need the
dominant mode of the histogram, the number of distinct elements in the sample,
and the total number of elements. In sample-based estimations, in the
worst-case scenario, the histogram might be as large as the view size we are
trying to estimate. Moreover, it is difficult to derive unassuming accuracy
bounds since the sample might not be representative. However, a sample-based
algorithm is expected to be an order of magnitude faster than an algorithm
which processes the entire data set.

Probabilistic counting~\cite{flajolet1985pca} and
 \textsc{LogLog} probabilistic counting (henceforth \textsc{LogLog})~\cite{durand2003lcl} have been
shown to provide very accurate unassuming view-size estimations quickly,
but their estimates assume we have independent hashing. Because of this
assumption, their theoretical bound may not hold in practice. Whether
this is a problem in practice is one of the contribution of this paper.

Gibbons and Tirthapura~\cite{Gibbons2001} derived an unassuming bound
(henceforth \textsc{Gibbons-Tirthapura})
that only requires pairwise independent hashing. It has been shown
recently that if you have $k$-wise independent hashing for $k>2$ the
theoretically bound can be improved substantially~\cite{viewsizetechreport}.
The benefit of \textsc{Gibbons-Tirthapura} is that as  long as the random number generator
is truly  random, the theoretical bounds have to hold irrespective of
the size of the view or of other factors.

All unassuming estimation techniques in this paper (\textsc{LogLog},
probabilistic counting and \textsc{Gibbons-Tirthapura}), have an accuracy proportional
to $1/\sqrt{M}$ where $M$ is a parameter noting the memory usage.

\section{\uppercase{Estimation by Multifractals}}\label{sec:multi}

\noindent We implemented the statistically assuming algorithm by Faloutsos~et~al. based on a multifractal model~\cite{faloutsos1996msd}.
Nadeau and Teorey~\cite{nadeau2003pmo}
reported competitive results for this approach.
Maybe surprisingly, given a sample, all that is required to learn the multifractal model
is the number of distinct elements in the sample $F_0$, the number of elements
in the sample $N'$, the total number of elements $N$, and the number of occurrences
of the most frequent item in the sample $m_\textrm{max}$.
Hence, a very simple
implementation is possible (see Algorithm~\ref{algo:multifractal}).
Faloutsos et al. erroneously introduced a tolerance factor $\epsilon$ in their algorithm:
unlike what they suggest, it is not possible, unfortunately, to adjust the model
parameter for an arbitrary
good fit, but instead, we have to be content with the best possible fit (see line~9
and following).

\begin{algorithm}
 \begin{small}\begin{algorithmic}[1]
\STATE \textbf{INPUT:} Fact table $t$ containing $N$ facts \STATE
\textbf{INPUT:} \textsc{group by} query on dimensions $D_1, D_2, \ldots, D_d$
\STATE \textbf{INPUT:} Sampling ratio $0<p<1$ \STATE \textbf{OUTPUT:} Estimated
size of \textsc{group by} query \STATE Choose a sample in $t'$ of size
$N'=\lfloor pN \rfloor$ \STATE Compute $g$=\textsc{group by}($t'$) \STATE let
$m_{\textrm{max}}$ be the number of occurrences of the most frequent tuple
$x_1,\ldots, x_d$ in $g$ \STATE let $F_0$ be the number of tuples in $g$ \STATE
$k \leftarrow \lceil\log  F_0 \rceil$ \WHILE{$F<F_0$} \STATE $p\leftarrow
(m_\textrm{max}/{N'})^{1/k}$ \STATE $F\leftarrow \sum_{a=0}^k {k\choose a}
(1-(p^{k-a}(1-p)^a)^{N'})$ \STATE $k \leftarrow k+1$ \ENDWHILE \STATE
$p\leftarrow (m_\textrm{max}/N)^{1/k}$ \STATE \textbf{RETURN: $\sum_{a=0}^k
{k\choose a} (1-(p^{k-a}(1-p)^a)^N)$}
 \end{algorithmic}
\end{small}
\caption{\label{algo:multifractal}View-size estimation using a multifractal distribution model.}
\end{algorithm}

\section{\uppercase{Unassuming View-Size Estimation}}

\subsection{Independent Hashing}\label{sec:hashing}

\noindent Hashing maps objects to values in a nearly random way. It has been
used for efficient data structures such as hash tables and in cryptography. We are interested in
hashing functions from tuples to $[0,2^L)$ where $L$ is fixed ($L=32$ in this
paper).  Hashing is uniform if $P(h(x)=y)=1/2^L$ for all $x,y$, that is, if all
hashed values are equally likely. Hashing is \textit{pairwise independent} if
$P(h(x_1)=y \land h(x_2)=z)= P(h(x_1)=y) P(h(x_2)=z)=1/4^L$ for all
$x_1,x_2,y,z$. Pairwise independence implies uniformity. Hashing is $k$-wise
independent if $P(h(x_1)=y_1 \land \cdots \land h(x_k)=y_k)=1/2^{kL}$ for all
$x_i,y_i$. Finally, hashing is (fully) independent if it is $k$-wise
independent for all $k$. It is believed that  independent hashing is unlikely
to be possible over large data sets using a small amount of
memory~\cite{durand2003lcl}.

Next, we show how 3-wise independent hashing is easily achieved in a
multidimensional data warehousing setting. For each dimension $D_i$, we build a
lookup table $T_{i}$, using the attribute values of $D_i$ as keys.
 Each time we meet a new key,
we generate a random number in $[0, 2L)$ and store it in the lookup table $T_i$.
This random number is the hashed value of this key.
 This table generates (fully)
independent hash values in amortized constant time. %This approach
%is reminiscent of bitmap indexes.
In a data warehousing context, whereas dimensions are
numerous, each dimension will typically have few distinct values: for example,
there are only 8,760 hours in a year. Therefore, the lookup table will often use a
few Mib or less. When hashing a tuple $x_1,x_2,\ldots,x_k$ in $D_1\times
D_2\times \ldots D_k$, we use the value $T_1(x_1) \lxor T_2(x_2) \lxor \cdots
\lxor T_k(x_k)$ where $ \lxor$ is the \textsc{exclusive or} operator. This hashing
 is 3-wise independent and requires amortized constant time.
 Tables $T_i$ can be reused for several estimations.

\subsection{Probabilistic Counting}\label{sec:probacounting}

\noindent Our implementation of (stochastic) probabilistic counting ~\cite{flajolet1985pca}
is given in  Algorithm~\ref{algo:stoch}. Recently, a variant of this algorithm,
\textsc{LogLog}, was proposed~\cite{durand2003lcl}.
Assuming independent hashing, these algorithms have standard error 
(or the standard deviation of the error) of
$0.78/\sqrt{M}$ and $1.3/\sqrt{M}$ respectively\danielcut{ (see
Figure~\ref{theorycounting})}. These theoretical results assume independent
hashing which we cannot realistically provide. Thus, we do not expect these
theoretical results to be always reliable.

\danielcut{
\begin{figure}
\begin{center}
\includegraphics[width=0.6\columnwidth,angle=270]{\myfig{loglog-stddev-vs-M}}
\end{center}
\caption{ \label{theorycounting}\daniel{This figure should probably
go since we did not correctly compute the standard error, but
rather the average error.}Standard error for probabilistic counting and
\textsc{LogLog}  as a function of the memory parameter
$M\in [128,2048]$: for most values of $M> 128$, the standard error is under
10\% for both of these algorithms. }
\end{figure}
}

\begin{algorithm}
\begin{small} \begin{algorithmic}[1]
\STATE \textbf{INPUT:} Fact table $t$ containing $N$ facts \STATE
\textbf{INPUT:} \textsc{group by} query on dimensions $D_1, D_2, \ldots, D_d$
\STATE \textbf{INPUT:} Memory budget parameter $M=2^k$ \STATE \textbf{INPUT:}
Independent hash function $h$ from $d$ tuples to $[0,2^L)$. \STATE
\textbf{OUTPUT:} Estimated size of \textsc{group by} query \STATE $b\leftarrow$
$M\times L$ matrix (initialized at zero) \FOR{tuple $x\in t$} \STATE
$x'\leftarrow \pi_{D_1, D_2, \ldots, D_d} (x)$ \COMMENT{projection of the
tuple} \STATE $y\leftarrow h(x')$ \COMMENT{hash $x'$ to $[0,2^L)$} \STATE
$\alpha = y \bmod{M}$ \STATE $i\leftarrow$ position of the first 1-bit in
$\lfloor y/M\rfloor$ \STATE $b_{\alpha,i}\leftarrow 1$ \ENDFOR \STATE $A
\leftarrow 0$ \FOR{$\alpha \in \{0,1,\ldots, M-1\}$} \STATE increment $A$ by
the position of the first zero-bit in $b_{\alpha,0},b_{\alpha,1},\ldots$
\ENDFOR \STATE \textbf{RETURN:} $M/\phi 2^{A/M}$ where $\phi\approx  0.77351$
 \end{algorithmic}\end{small}

\caption{\label{algo:stoch}View-size estimation using (stochastic) probabilistic counting.}
\end{algorithm}

\begin{algorithm}
\begin{small} \begin{algorithmic}[1]
\STATE \textbf{INPUT:} fact table $t$ containing $N$ facts
\STATE \textbf{INPUT:} \textsc{group by} query on dimensions $D_1, D_2, \ldots, D_d$
\STATE \textbf{INPUT:} Memory budget parameter $M=2^k$
\STATE \textbf{INPUT:} Independent hash function $h$ from $d$ tuples to $[0,2^L)$.
\STATE \textbf{OUTPUT:} Estimated size of \textsc{group by} query
\STATE $\mathcal{M}\leftarrow \underbrace{0,0,\ldots,0}_M$
\FOR{tuple $x\in t$}
\STATE $x'\leftarrow \pi_{D_1, D_2, \ldots, D_d} (x)$ \COMMENT{projection of the tuple}
\STATE $y\leftarrow h(x')$ \COMMENT{hash $x'$ to $[0,2^L)$}
\STATE $j\leftarrow$ value of the first $k$ bits of $y$ in base 2
\STATE $z\leftarrow$ position of the first 1-bit in the remaining $L-k$ bits of $y$ (count starts at 1)
\STATE $\mathcal{M}_j \leftarrow \max (\mathcal{M}_j,z)$
\ENDFOR
\STATE \textbf{RETURN:} $\alpha_M M 2^{\frac{1}{M}\sum_j \mathcal{M}_j}$ where $\alpha_M  \approx 0.39701-(2\pi^2+\ln^2 2)/(48M)$.
 \end{algorithmic}\end{small}

\caption{\label{algo:loglog}View-size estimation using \textsc{LogLog}.}
\end{algorithm}

\subsection{\textsc{Gibbons-Tirthapura}}\label{sec:gibbons}

\noindent Our implementation of the  \textsc{Gibbons-Tirthapura} algorithm (see
Algorithm~\ref{algo:gibbons}) hashes each tuple only once unlike the original
algorithm~\cite{Gibbons2001}. Moreover, the independence of the hashing depends
on the number of dimensions used by the \textsc{group by}. If the view-size is
smaller than the memory parameter ($M$), the view-size estimation is without
error. For this reason, we expect \textsc{Gibbons-Tirthapura} to perform well
when estimating
small and moderate view sizes. %We assume that $L$ is such that $ 2^L M=2^{32} M$
%is larger than the view-size we estimate.

\begin{algorithm}
\begin{small} \begin{algorithmic}[1]
\STATE \textbf{INPUT:} Fact table $t$ containing $N$ facts \STATE
\textbf{INPUT:} \textsc{group by} query on dimensions $D_1, D_2, \ldots, D_d$
\STATE \textbf{INPUT:} Memory budget parameter $M$ \STATE \textbf{INPUT:}
$k$-wise  hash function $h$ from $d$ tuples to $[0,2^L)$. \STATE
\textbf{OUTPUT:} Estimated size of \textsc{group by} query \STATE
$\mathcal{M}\leftarrow$ empty lookup table \STATE $t\leftarrow0$ \FOR{tuple
$x\in t$} \STATE $x'\leftarrow \pi_{D_1, D_2, \ldots, D_d} (x)$
\COMMENT{projection of the tuple} \STATE $y\leftarrow h(x')$ \COMMENT{hash $x'$
to $[0,2^L)$} \STATE $j\leftarrow$ position of the first 1-bit in $y$ (count
starts at 0) \IF{$j \leq t$} \STATE $\mathcal{M}_{x'}=j$
\WHILE{$\textrm{size}(\mathcal{M})>M$} \STATE $t\leftarrow t+1$ \STATE prune
all entries in $\mathcal{M}$ having value less than $t$ \ENDWHILE \ENDIF
\ENDFOR \STATE \textbf{RETURN:}.$2^t \textrm{size}(\mathcal{M})$
 \end{algorithmic}\end{small}
\caption{\label{algo:gibbons}\textsc{Gibbons-Tirthapura} view-size estimation.}
\end{algorithm}

\begin{figure}
\begin{center}
\includegraphics[width=0.6\columnwidth,angle=270]{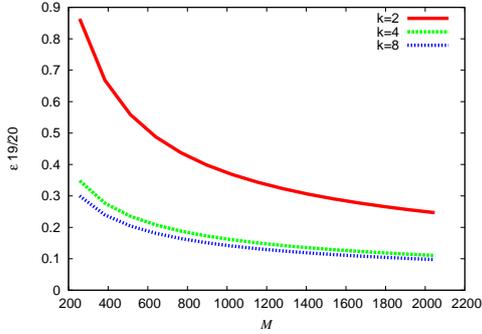}
\end{center}
\caption{
\label{theory}Bound on the estimation error (19 times out of 20) as a function
of the number of tuples kept in memory ($M\in [128,2048]$) according to Proposition~\ref{countprop}
for \textsc{Gibbons-Tirthapura} view-size estimation with $k$-wise independent hashing. %We prove that at least 2~comparisons per element are required when no stream latency is allowed.
}
\end{figure}

The theoretical bounds given in~\cite{Gibbons2001} assumed
pairwise independence. The generalization below is from~\cite{viewsizetechreport} and is illustrated by Figure~\ref{theory}.

\begin{proposition}\label{countprop}Algorithm~\ref{algo:gibbons} estimates the number of distinct tuples
within relative precision $\epsilon$,
with a  $k$-wise independent hash for $k\geq 2$ by storing $M$ distinct tuples ($M\geq 8 k$)
and  with reliability $1-\delta$ where $\delta$ is given by
\[\delta= \frac{k^{k/2}}{ e^{k/3}  M^{k/2}} \left (  2^{k/2}
+\frac{ 8^{k/2}}{ \epsilon^k  (2^{k/2}-1)} \right ).\]
More generally, we have
\newcommand{\frack}[2]{{{#1}/{#2}}}  % ensure similar formatting in the general
\begin{eqnarray*}\delta
  &\leq & \frac{k^\frack{k}{2}}{ e^{\frack{k}{3}}  M^{\frack{k}{2}}  } \left ( \frac{\alpha^{\frack{k}{2}}}{(1-\alpha)^k}+ \frac{4^{\frack{k}{2}}}{\alpha^{\frack{k}{2}} \epsilon^k (2^{\frack{k}{2}}-1)}\right ) .
\end{eqnarray*}
for $4k/M\leq \alpha<1$ and any $k,M>0$.
\end{proposition}

In the case where hashing is 4-wise independent, 
we derive a more concise bound.

\begin{corollary}\label{corollary1}
 With 4-wise independent hashing, Algorithm~\ref{algo:gibbons} estimates
 the number of distinct tuples
within relative precision $\epsilon\approx 5/\sqrt{M} $, 19 times out of 20
for $\epsilon$ small.
\end{corollary}
\begin{proof}\newcommand{\frack}[2]{{{#1}/{#2}}}  % ensure similar formatting in the general
 We start from the second inequality of Proposition~\ref{countprop}.
Differentiating $\frac{\alpha^{\frack{k}{2}}}{(1-\alpha)^k}+\frac{4^{\frack{k}{2}}}{\alpha^{\frack{k}{2}} \epsilon^k (2^{\frack{k}{2}}-1)}$
with respect to $\alpha$ and setting the result to zero, we get
$3\alpha^4 \epsilon^4+16 \alpha^3-48 \alpha^2-16=0$ (recall that $4k/M\leq \alpha<1$).
By multiscale analysis, we seek a solution of the form $\alpha = 1-a\epsilon^r+o(\epsilon^r)$ and we have that $\alpha\approx 1- \frack{1}{2}\sqrt[3]{\frack{3}{2}} \epsilon^{4/3}$ for $\epsilon$ small. Substituting this value of $\alpha$, we have $\frac{\alpha^{\frack{k}{2}}}{(1-\alpha)^k}+ \frac{4^{\frack{k}{2}}}{\alpha^{\frack{k}{2}} \epsilon^k (2^{\frack{k}{2}}-1)}\approx \frack{128}{24 \epsilon^4}$. The result follows by substituting in the second inequality.
\end{proof}

\section{\uppercase{Experimental Results}}\label{sec:Experiment}

\noindent To benchmark the quality of the view-size estimation against the memory
and speed, we have run test over the US~Census~1990 data
set~\cite{KDDRepository} as well as on synthetic data produced by
DBGEN~\cite{DBGEN}. The synthetic data was produced by running the DBGEN
application with scale factor parameter equal to 2. The characteristics of
data sets are detailed in Table~\ref{tab:caractDataSet}. We selected
20 and 8 views respectively from these data sets: all views in US~Census~1990
have at least 4~dimensions whereas only 2 views have at least 4~dimensions in the synthetic
data set.

%Moreover, we computed the exact size of 20 views ranging within $[389,1816063]$
%for US~Census~1990 data
%set and 8 views ranging within $[376643,13997981]$ for DBGEN data set.
%(May be it is suitable to put those information in a table)

\begin{table}
    \centering
    \begin{tabular}{ccc} \hline
     & \textbf{US~Census~1990} & \textbf{DBGEN} \\ \hline
    \# of facts& 2458285 & 13977981 \\
    \# of views& 20 & 8 \\
    %View size interval& $[389,1816063]$ & $[376643,13997981]$ \\ \hline
    \# of attributes& 69 & 16 \\ \hline
    Data size& 360\,MB & 1.5\,GB\\ \hline
    \end{tabular}
    \caption{Characteristic of data sets.}\label{tab:caractDataSet}
\end{table}

We used the GNU C++ compiler version~4.0.2 with the ``-O2'' optimization flag on a Centrino Duo 1.83\,GHz machine with  2\,GB of
RAM running Linux kernel 2.6.13--15. No thrashing was observed. To ensure
reproducibility, C++ source code is available freely from the authors.

For the US~Census~1990 data set, the hashing look-up table is a simple array
since there are always fewer than 100~attribute values per dimension.
Otherwise, for the synthetic DBGEN data, we used the GNU/CGI STL extension
\texttt{hash\_map} which is to be integrated in the C++ standard as an
\texttt{unordered\_map}: it provides amortized $O(1)$ inserts and queries. All
other look-up tables are implemented using the STL \texttt{map} template which
has the same performance characteristics of a red-black tree. We used comma
separated (CSV) (and pipe separated files for DBGEN) text files and wrote our
own C++ parsing code.
%
% Discuss here the number of dimensions.

The test protocol we adopted (see Algorithm~\ref{algo:protocol}) has been
executed for each estimation technique (\textsc{LogLog}, probabilistic counting
and \textsc{Gibbons-Tirthapura}), \textsc{group by} query, random seed and
memory size. At each step corresponding to those parameter values, we compute
the estimated-size values of \textsc{GROUP BY}s and time required for their
computation.
%We also computed the exact values of \textsc{GROUP BYs} in order to derive the
%view-size estimation accuracy (error).
For the multifractal estimation technique, we computed at the same way the time
and estimated size for each \textsc{GROUP BY}, sampling ratio value and random
seed.

\begin{figure*}[htb]
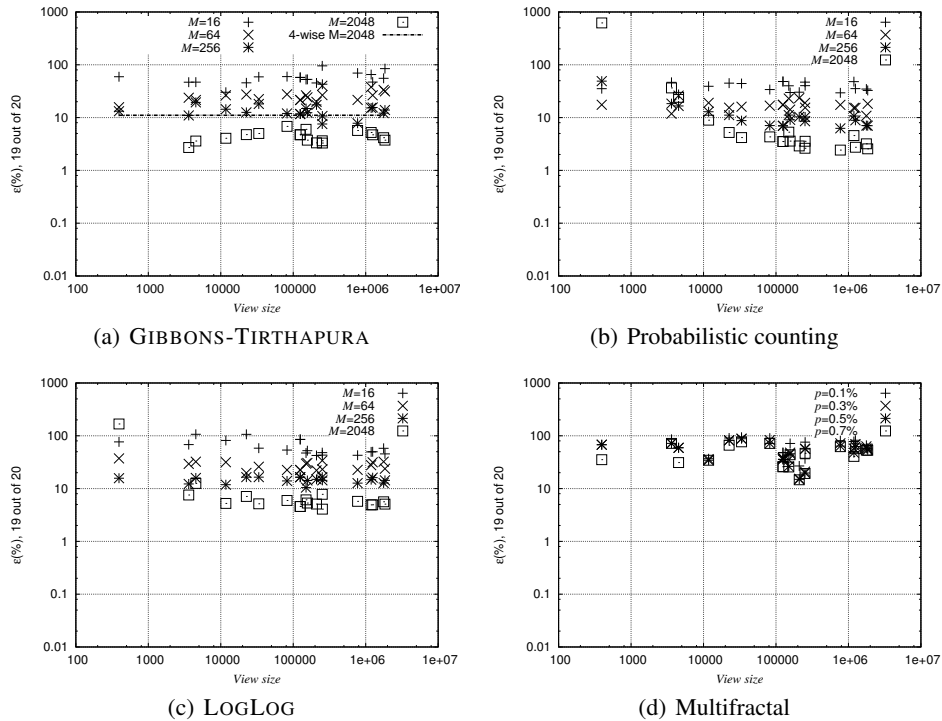

\begin{center}
  \subfigure[\textsc{Gibbons-Tirthapura}]{\includegraphics[width=0.8\columnwidth]{\myfig{gibbons_error_19out20}}\label{fig:expUscensus19Out20Gibbons}}\quad
  \subfigure[Probabilistic counting]{\includegraphics[width=0.8\columnwidth]{\myfig{counting_error_19out20}}\label{fig:expUscensus19Out20Counting}}
%\danielcut{
\\ [1pt]
  \subfigure[\textsc{LogLog}]{\includegraphics[width=0.8\columnwidth]{\myfig{loglog_error_19out20}}\label{fig:expUscensus19Out20LogLog}}\quad
  \subfigure[Multifractal]{\includegraphics[width=0.8\columnwidth]{\myfig{multifractal_error_19out20}}\label{fig:expUscensus19Out20Multi}}%}
\end{center}
\caption{$95^{\textrm{th}}$-percentile error 19/20 $\epsilon$ as a function of
exact view size for increasing values of $M$ (US Census 1990)} \label{fig:expUscensus19Out20}
\end{figure*}

%In the following sections, we present the results obtained for each data set.
%\danielcut{
\begin{algorithm}
\begin{small}\begin{algorithmic}[1]
%\STATE \textbf{INPUT:} Fact table $F$ containing $N$ facts

%\STATE \textbf{INPUT:} \textsc{group by} queries on dimensions $Q$

%\STATE \textbf{INPUT:} Memory budget values $M$

%\STATE \textbf{INPUT:} Random seed values $R$

%\STATE \textbf{OUTPUT:} Log file in which estimation results are stored

\FOR{\textsc{group by} query $q\in Q$}

    \FOR{memory budget $m\in M$}

        \FOR{random seed value $r\in R$}

            \STATE Estimate the size of \textsc{group by} $q$ with $m$ memory budget and $r$
            random seed value
            \STATE Save estimation results (time and estimated size) in a log
            file
        \ENDFOR
    \ENDFOR
\ENDFOR

\end{algorithmic}\end{small}
\caption{\label{algo:protocol}Test protocol.}
\end{algorithm}
%}
%\subsection{Accuracy of estimates}
%}
\noindent  \textbf{US Census 1990.} Figure~\ref{fig:expUscensus19Out20} plots the largest
$95^{\textrm{th}}$-percentile error observed over 20 test estimations for
various memory size $M \in \{16, 64, 256, 2048\}$. For the
multifractal estimation technique, we represent the error for each sampling
ratio $p \in \{0.1\%, 0.3\%, 0.5\%, 0.7\%\}$.  The X axis represents the size of the
exact \textsc{GROUP BY} values. This
$95^{\textrm{th}}$-percentile error can be related to the theoretical bound for
$\epsilon$ with 19/20 reliability for \textsc{Gibbons-Tirthapura} (see
Corollary~\ref{corollary1}): we see that this upper bound is verified experimentally.
However, the error on ``small'' view sizes can exceed 100\% for probabilistic counting
and \textsc{LogLog}.

\noindent \textbf{Synthetic data set.}  Similarly, we computed \danielcut{the standard
error\daniel{NO!!! average error} and} the 19/20 error for each technique, computed from the DDBGEN data set
\danielcut{(see Figure~\ref{fig:expDbgenError} and Figure~\ref{fig:expDbgen19Out20})}.
We observed that the four techniques have the same behaviour observed on the US
Census data set. Only, this time, the theoretical bound for the 19/20 error is larger
\danielcut{in Figure~\ref{fig:expDbgen19Out20Gibbons} than it was in Figure~\ref{fig:expUscensus19Out20Gibbons}}
because the synthetic data sets has many views with less than 2~dimensions\danielcut{ and
so the hashing is no more than pairwise independent}.

\noindent \textbf{Speed.} We have also computed the time needed for each
technique to estimate view-sizes. We do not represent this time because it is
similar for each technique except for the multifractal which is the fastest
one. In addition, we observed that time do not depend on the memory budget
because most time is spent streaming and hashing the data.
 For the multifractal technique, the processing time increases with the
sampling ratio.

The time needed to estimate the size of all the views by \textsc{Gibbons-Tirthapura}, probabilistic counting and
\textsc{LogLog} is about 5 minutes for US~Census~1990 data set and 7 minutes for the synthetic
data set. For the multifractal technique, all the estimates are done on roughly
2 seconds. This time does not include the time needed for sampling data which can
be significant: it takes 1~minute (resp. 4~minutes) to sample  0.5\%
of the US Census data set (resp. the synthetic data set -- TPC~H)
because the data is not stored in a flat file.

\section{\uppercase{Discussion}}

\noindent  Our results show that  probabilistic counting and \textsc{LogLog} do
not entirely live up to their  theoretical promise. For small view sizes, the
relative accuracy can be very low.

 When comparing the memory usage of the various techniques,
we have to keep in mind that the  memory parameter $M$
can translate in different memory usage. The memory usage depends
also on the number of dimensions of each view. Generally, \textsc{Gibbons-Tirthapura}
will use more memory for the same value of $M$ than either probabilistic counting or
\textsc{LogLog}, though all of these can be small compared to the memory
usage of the lookup tables $T_i$ used for 3-wise independent hashing.
In this paper, the memory usage was always of the order of a few MiB which
is negligible in a data warehousing context.

View-size estimation by sampling can take minutes when data is not layed out
in a flat file or indexed, but
the time required for an unassuming estimation is even higher.
Streaming and hashing the tuples accounts for most of the processing time so
for faster estimates, we could store all hashed values in a bitmap (one per dimension).

% If accuracy matters, either probabilistic counting or  \textsc{Gibbons-Tirthapura}
% are the best choices with  \textsc{Gibbons-Tirthapura} being better choice when
% reliability is important.  If only ballpark figures
% are required, estimation by multifractals is adequate.

\section{\uppercase{Conclusion and future work}}\label{sec:Conclusion}

\noindent In this paper, we have provided unassuming techniques for view-size
estimation in a data warehousing context. We adapted an estimator due to
Gibbons and Tirthapura. We compared this technique experimentally with
stochastic probabilistic counting, \textsc{LogLog}, and
multifractal statistical models. We have demonstrated that among these
techniques, only \textsc{Gibbons-Tirthapura} provides stable estimates irrespective of
the size of views. Otherwise, (stochastic) probabilistic counting  has  a small edge
in accuracy for relatively large views, whereas the competitive sampling-based
technique (multifractal) \danielcut{we tested} is an order of magnitude faster but can
\danielcut{sometimes} provide crude estimates. According to our experiments, \textsc{LogLog} was not faster than either \textsc{Gibbons-Tirthapura}
or probabilistic counting, and since it is less accurate than probabilistic counting, we
cannot recommend it.
%\danielcut{
There is ample room for future work. Firstly,
we plan to extend
these techniques to other types of aggregated views (for
example, views including \textsc{Having} clauses). Secondly,
we want to precompute the hashed values for very fast view-size estimation.
 Furthermore, these
techniques should be tested in a  materialized view selection
heuristic.
%}

%what should one be using?

%Safe: Gibbons
%Accurate: probabilistic counting
%Fast: stat.

%we can't recommend \textsc{LogLog} probabilistic counting

\section*{\uppercase{Acknowledgements}}

\noindent The authors wish to thank Owen Kaser for hardware and software.
This work is supported by NSERC grant 261437 and by  FQRNT grant 112381.

%\vfill

\renewcommand{\baselinestretch}{0.5}%98
\bibliographystyle{apalike}
{\small
%\singlespace
%\balance
\bibliography{../bib/lemur}}
\renewcommand{\baselinestretch}{1}

\end{document}